\documentclass[a4paper,twocolumn,11pt,accepted=2022-07-02]{quantumarticle}
\pdfoutput=1
\usepackage[utf8]{inputenc}
\usepackage[english]{babel}
\usepackage[T1]{fontenc}
\usepackage{amsmath}
\usepackage{hyperref}
\usepackage{tikz}
\usepackage{lipsum}

\usepackage{bm}
\usepackage{amssymb}
\usepackage{amsfonts}
\usepackage{latexsym}
\usepackage{enumerate}
\usepackage{algorithm}
\usepackage{algorithmic}
\usepackage{cases}

\newcommand{\white}{\color{white}}
\newtheorem{theorem}{Theorem}
\newtheorem{definition}{Definition}
\makeatletter
\renewcommand*{\@opargbegintheorem}[3]{\trivlist
      \item[\hskip \labelsep{\bfseries #1\ #2}] \textbf{(#3)}\ \itshape}
\makeatother

\begin{document}

\title{Divide-and-conquer verification method for noisy intermediate-scale quantum computation}

\author{Yuki Takeuchi}
\affiliation{NTT Communication Science Laboratories, NTT Corporation, 3-1 Morinosato Wakamiya, Atsugi, Kanagawa 243-0198, Japan}
\orcid{0000-0003-2428-7432}
\email{yuki.takeuchi.yt@hco.ntt.co.jp}
\author{Yasuhiro Takahashi}
\affiliation{NTT Communication Science Laboratories, NTT Corporation, 3-1 Morinosato Wakamiya, Atsugi, Kanagawa 243-0198, Japan}
\affiliation{Faculty of Informatics, Gunma University, 4-2 Aramakimachi, Maebashi, Gunma 371-8510, Japan}
\author{Tomoyuki Morimae}
\affiliation{Yukawa Institute for Theoretical Physics, Kyoto University, Kitashirakawa Oiwakecho, Sakyo-ku, Kyoto 606-8502, Japan}
\author{Seiichiro Tani}
\affiliation{NTT Communication Science Laboratories, NTT Corporation, 3-1 Morinosato Wakamiya, Atsugi, Kanagawa 243-0198, Japan}
\affiliation{International Research Frontiers Initiative (IRFI), Tokyo Institute of Technology, Japan}
\orcid{0000-0002-6041-1704}
\maketitle

\begin{abstract}
Several noisy intermediate-scale quantum computations can be regarded as logarithmic-depth quantum circuits on a sparse quantum computing chip, where two-qubit gates can be directly applied on only some pairs of qubits.
In this paper, we propose a method to efficiently verify such noisy intermediate-scale quantum computation.
To this end, we first characterize small-scale quantum operations with respect to the diamond norm.
Then by using these characterized quantum operations, we estimate the fidelity $\langle\psi_t|\hat{\rho}_{\rm out}|\psi_t\rangle$ between an actual $n$-qubit output state $\hat{\rho}_{\rm out}$ obtained from the noisy intermediate-scale quantum computation and the ideal output state (i.e., the target state) $|\psi_t\rangle$.
Although the direct fidelity estimation method requires $O(2^n)$ copies of $\hat{\rho}_{\rm out}$ on average, our method requires only $O(D^32^{12D})$ copies even in the worst case, where $D$ is the denseness of $|\psi_t\rangle$. For logarithmic-depth quantum circuits on a sparse chip, $D$ is at most $O(\log{n})$, and thus $O(D^32^{12D})$ is a polynomial in $n$.
By using the IBM Manila 5-qubit chip, we also perform a proof-of-principle experiment to observe the practical performance of our method.
\end{abstract}

\section{Introduction}
Universal quantum computers are expected to efficiently solve several hard problems that are intractable for classical counterparts.
However, to exploit their full potential, quantum error correction (QEC) is necessary.
For current technologies, QEC is highly demanding because it requires precise state preparations, quantum operations, and measurements.
That is why the potential of quantum computation without QEC is being actively explored.
Such non-fault-tolerant quantum computation is called noisy intermediate-scale quantum (NISQ) computation~\cite{P18}, and several quantum algorithms tailored for it have already been proposed~\cite{PMSYZLAO14,FGG14,MNKF18} and experimentally evaluated~\cite{KMTTBCG17,HCTHKCG19}.

\begin{figure}[t]
\centering
\includegraphics[width=8cm, clip]{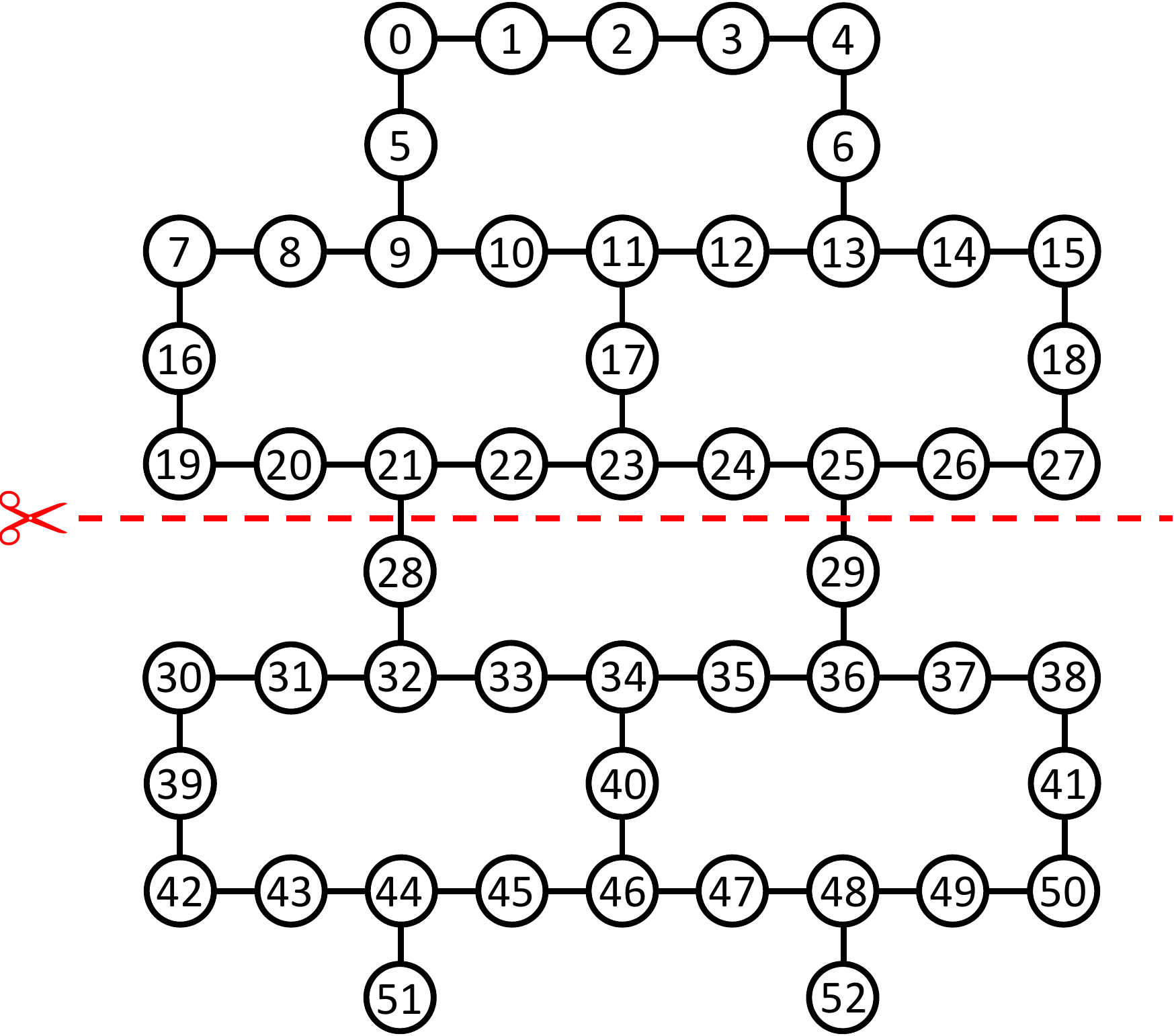}
\caption{The connectivity of qubits within the IBM Rochester $53$-qubit chip. Two-qubit operations can be directly applied on only pairs of two qubits connected by a line.  This chip can be separated into two parts ($0-27$ and $28-52$) by removing only two connections between the $21$st and $28$th qubits and between the $25$th and $29$th qubits.}
\label{IBMQ}
\end{figure}

Although several error mitigation techniques have been proposed~\cite{LB17,TBG17,EBL18,PJ18,MSSO18,SYTVBE20,ZKFYGW20,SQCBL20,CACC20,ZG20}, NISQ computation should be finished in at most logarithmic time due to the lack of fault-tolerance.
Note that when an error occurs with a constant probability in each time step, logarithmic-depth quantum circuits succeed with a probability of the inverse of a polynomial.
Furthermore, some current quantum computing chips are sparse in the sense that they can be separated into two parts by removing a small number of connections between two qubits.
For example, IBM's 53-qubit chip~\cite{APS19,TC20,PPMNGPVL20,K20,DHB20,SSM20,W20,HCCHHC20} in Fig.~\ref{IBMQ} can be separated into two parts ($0-27$ and $28-52$) by removing only two connections between the 21st and 28th qubits and between the 25th and 29th qubits.
In short, several NISQ computations can be regarded as shallow (i.e., at most logarithmic-depth) quantum circuits on a sparse chip, and we focus on such NISQ computations.

Since the performance of NISQ computations is strongly affected by noise, it is necessary to efficiently check whether a given NISQ computer works as expected.
This task is known as the verification of quantum computation.
Although various efficient verification methods~\cite{M14,HM15,M16,AOEM17,FK17,MTH17,FHM18,TM18,B18,M18,TMMMF19,HT19,GV19,ACGH19,ZH19,CCY19,MK20} have been proposed so far, they all assume (fault-tolerant) universal quantum computations.
Particularly, some of these methods~\cite{M14,HM15,M16,FK17,MTH17,TM18,TMMMF19,HT19,ZH19,MK20} are based on measurement-based quantum computation (MBQC)~\cite{RB01}, which consumes at least one qubit to apply a single elementary quantum gate.
Therefore, MBQC requires more qubits than the quantum circuit model.
Since the number of available qubits is limited in NISQ computations, the MBQC is inadequate for it.
The method of Fitzsimons {\it et al.}~\cite{FHM18} requires a prover (i.e., a quantum computer to be verified) to generate a Feynman-Kitaev history state whose generation seems to be hard for NISQ computations.
Other methods~\cite{M18,GV19,ACGH19,CCY19} require the prover to compute classical functions in a superposition, where the functions are constructed from the learning-with-errors problem~\cite{R05}.
This task also seems to be beyond the capability of NISQ computations.
Furthermore, since NISQ computers are expected to be used to solve several problems such as optimizations, classifications, and simulations of materials, a verification method should be developed for general problems other than decision problems, which can be answered by YES or NO.

\begin{figure*}[t]
\centering
\includegraphics[width=13cm, clip]{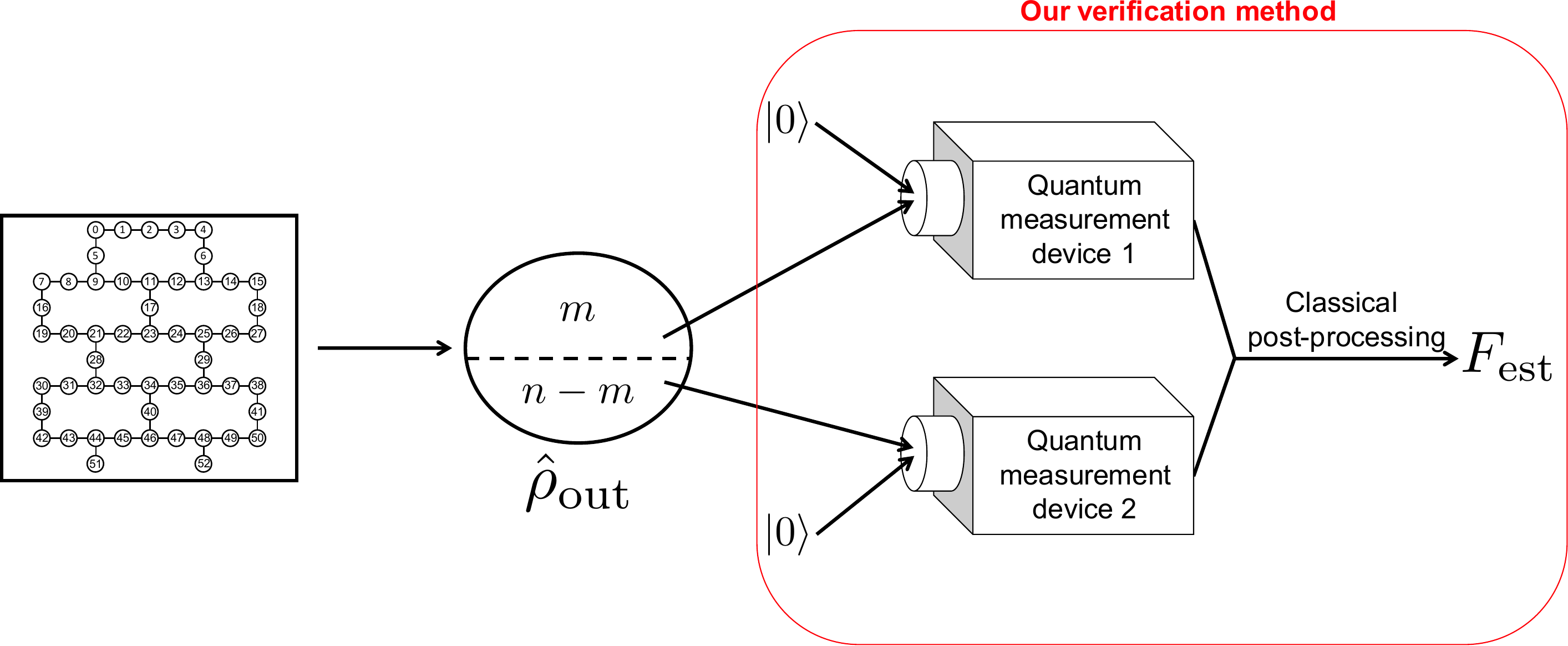}
\caption{Diagram of our verification method. The $n$-qubit quantum state $\hat{\rho}_{\rm out}$ is the output from a sparse chip. $m$ and other $n-m$ qubits of $\hat{\rho}_{\rm out}$ are measured by noisy quantum measurement devices 1 and 2 with an ancillary qubit $|0\rangle$, respectively. Our method repeats these procedures polynomially many times. Then by classically post-processing all measurement outcomes, we obtain an estimate $F_{\rm est}$ of the fidelity $\langle\psi_t|\hat{\rho}_{\rm out}|\psi_t\rangle$. Although we depict the quantum measurement devices 1 and 2 as different devices for simplicity, they can be treated as a single $(m+1)$-qubit device by sequentially measuring $m$ and the other $n-m$ qubits of $\hat{\rho}_{\rm out}$.}
\label{diagram}
\end{figure*}

It is highly nontrivial whether an $n$-qubit NISQ computer can be efficiently verified by using a quantum measurement device that is strictly smaller than $n$ qubits~\cite{FN+}.
In this paper, we solve this problem affirmatively.
More precisely, let $|\psi_t\rangle$ be any target state, i.e., any $n$-qubit pure state generated from an ideal logarithmic-depth quantum circuit on a sparse chip.
We propose an efficient method (Algorithm~1) to estimate the fidelity between $|\psi_t\rangle$ and the actual state $\hat{\rho}_{\rm out}$~\cite{FN} generated from an actual NISQ computer. Our method needs a $(m+1)$-qubit measurement device, where $n/2+1\le m+1<n$.
Since our method estimates the fidelity between the actual and ideal quantum states, it can be used for any problems beyond decision problems.
Our method is constructed as follows (see also Fig.~\ref{diagram}): first, we obtain an upper bound on the diamond norms between the ideal quantum operations achieved in the $(m+1)$-qubit measurement device and its actual ones.
Our method works even if the $(m+1)$-qubit measurement device is somewhat noisy, i.e., the upper bound is non-zero but sufficiently small.
Then we measure $m$ qubits of $\hat{\rho}_{\rm out}$ and an ancillary qubit $|0\rangle$ by using the noisy $(m+1)$-qubit operators.
The remaining $n-m$ qubits of $\hat{\rho}_{\rm out}$ are also similarly measured with another ancillary qubit $|0\rangle$, where $n-m\le m$.
We repeat these procedures by generating a polynomial number of copies of $\hat{\rho}_{\rm out}$.
Finally, by classically post-processing all measurement outcomes, we estimate the fidelity $\langle\psi_t|\hat{\rho}_{\rm out}|\psi_t\rangle$.
Since we divide $\hat{\rho}_{\rm out}$ into two parts and measure each of them separately, our verification method can be considered as a divide-and-conquer method.

Our verification method is superior to existing methods in terms of the number of required copies, which we call the sample complexity.
The sample complexity of our method is $O(D^32^{12D})$ that is a polynomial in $n$ for NISQ computations because the denseness $D$ of $|\psi_t\rangle$ is at most $O(\log{n})$ (for the definition of the denseness, see Definition~\ref{sparse2}).
As an existing fidelity estimation method, the quantum state tomography~\cite{SBRF93,H97,BDPS99} can estimate the fidelity by reconstructing the matrix representation of $\hat{\rho}_{\rm out}$.
Since any $n$-qubit state can be identified using $O(4^n)$ complex numbers, the quantum state tomography requires at least the same number of copies of $\hat{\rho}_{\rm out}$.
To improve the efficiency, Flammia and Liu proposed a direct fidelity estimation method that estimates the fidelity without reconstructing $\hat{\rho}_{\rm out}$~\cite{FL11}.
Their method requires $\Omega(2^n)$ copies of $\hat{\rho}_{\rm out}$ on average.
Although their method improved on the quantum state tomography in terms of the number of required copies, the number is still exponential in $n$.
On the other hand, as mentioned above, our method requires only a polynomial number of copies even in the worst case.

Prior to our work, as far as we know, only two types of verification methods were developed for NISQ computers~\cite{FKD19,FMMD21,LMKO21}.
Our method can estimate the fidelity between ideal and actual output states, unlike the methods in Refs.~\cite{FKD19,FMMD21}, which can estimate the total variation distance between an actual output probability distribution and the ideal one.
In this sense, our method is superior to theirs.
Note that theirs can also be used to estimate lower and upper bounds on the probability that the target quantum circuit is afflicted by errors.
By using the upper bound, a lower bound on the fidelity can be obtained, which may be loose.
The method in Refs.~\cite{LMKO21} achieves both verifiability and the security, i.e., it enables us to securely and verifiably delegate quantum computing to a remote server even if the server's quantum computer is noisy.
Its error robustness is promising.
However, their method is based on the MBQC, which seems to be inadequate for NISQ computers as we have mentioned, whereas our method is not.

Efficient verification methods have already been proposed for several quantum states such as graph states~\cite{HM15,MK20,TMMMF19}, hypergraph states~\cite{MTH17,TM18,ZH19}, weighted graph states~\cite{HT19}, and Dicke states~\cite{LYSZZ19}.
Since these previous methods are tailored for the specific classes of states, they cannot be used for our purpose.
Our method can efficiently estimate the fidelity for any pure state generated by shallow quantum circuits on a sparse chip.

The rest of this paper is organized as follows. In Sec.~\ref{II}, we formally define a class of quantum states, which will be proved to be efficiently verifiable with our method.
In Sec.~\ref{III}, we construct our verification method and derive its sample complexity.
In Sec.~\ref{addIV}, we provide an observation on the practical performance of our method by performing a proof-of-principle experiment on the IBM Manila 5-qubit chip.
Section~\ref{IV} is devoted to the conclusion and discussion.

\section{Target state: sparse quantum state}
\label{II}
In this section, we formally define a class of quantum states to be efficiently verified by our method.
More precisely, we define sparse quantum states as output states of shallow quantum circuits implemented on a sparse quantum computing chip.
To this end, we define a sparse quantum computing chip.
Quantum computing chips are often represented by graphs whose vertices and edges represent qubits and the connectivity between qubits (i.e., the applicability of two-qubit gates), respectively.
Two-qubit gates can be directly applied on two qubits whose corresponding vertices are connected by an edge.
By using this graph representation, sparse quantum computing chips are defined as follows:
\begin{definition}[Sparse quantum computing chip]
\label{sparse}
An $n$-qubit quantum computing chip ${\mathcal C}$ is a graph such that $n$ qubits are located on each vertex, and each edge represents the connectivity between the qubits (i.e., the applicability of two-qubit gates).
${\mathcal C}$ is called sparse if it can be divided into two subgraphs with $\Theta(n)$ vertices by removing a constant number of edges.
\end{definition}
As a simple example, when ${\mathcal C}$ is a one-dimensional graph, it is sparse.
This is because, a one-dimensional graph can be divided into two subgraphs by removing only one edge.

We also define the denseness of pure states as follows:
\begin{definition}[Denseness of pure states]
\label{sparse2}
Let us consider a gate set composed of all single-qubit gates and the controlled-$Z$ $(CZ)$ gate $|0\rangle\langle 0|\otimes I+|1\rangle\langle 1|\otimes Z$, where $I$ and $Z$ are the two-dimensional identity gate and the Pauli-$Z$ gate, respectively.
Let $U$ be any $n$-qubit quantum circuit that consists of a polynomial number of quantum gates chosen from the gate set.
When the set of the $n$ qubits are partitioned into two sets $A$ and $B$ of $\Theta(n)$ qubits, let $CUT(A:B)$ be the number of $CZ$ gates that act across $A$ and $B$ in the quantum circuit $U$.
For every such circuit $U$, the denseness $D$ of the pure state $U|0^n\rangle$ with respect to $U$ is the minimum number of $CUT(A:B)$ over all partitions $A:B$ of the set of the $n$ qubits that $U$ acts on.
When $U$ is clear from the context, we may just write ``the denseness $D$ of the pure state $|\psi\rangle$", where $|\psi\rangle=U|0^n\rangle$.
\end{definition}
From Definition~\ref{sparse}, when an $n$-qubit quantum circuit $U$ is a $d$-depth quantum circuit on a sparse chip, the denseness $D$ of $|\psi_t\rangle\equiv U|0^n\rangle$ is $O(d)$.
Therefore, when we consider NISQ computations satisfying $d=O(\log{n})$, the denseness $D$ is also $O(\log{n})$.

\section{Divide-and-conquer verification method}
\label{III}
In this section, we give our main result.
We explain our construction step by step.
In Sec.~\ref{IIIA}, we introduce generalized stabilizer operators and explain how to estimate the fidelity by using them.
In Sec.~\ref{IIIB}, we decompose $n$-qubit generalized stabilizer operators into linear combinations of tensor products of $m$-qubit and $(n-m)$-qubit operators.
Then we give an $(n+1)$-qubit quantum circuit to measure each term of the linear combinations.
In Sec.~\ref{IIIC}, we show that the $(n+1)$-qubit quantum circuit can be replaced with $(m+1)$-qubit and $(n-m+1)$-qubit quantum circuits.
This means that the fidelity $\langle\psi_t|\hat{\rho}_{\rm out}|\psi_t\rangle$ can be estimated using the resultant $(m+1)$-qubit and $(n-m+1)$-qubit quantum circuits.
We also summarize the procedures of our verification method as Algorithm~1.
Finally, in Sec.~\ref{IIID}, we derive the sample complexity, i.e., the required number of copies of $\hat{\rho}_{\rm out}$ to estimate the fidelity.
This sample complexity is summarized as Theorem~\ref{theorem}.

\subsection{Generalized stabilizer operators}
\label{IIIA}
Our final goal is to estimate the fidelity $\langle\psi_t|\hat{\rho}_{\rm out}|\psi_t\rangle$ with the ideal sparse quantum state $|\psi_t\rangle\equiv U|0^n\rangle$.
To this end, we first introduce generalized stabilizer operators $\{g_i\}_{i=1}^n$ for $|\psi_t\rangle$, where $g_i\equiv UZ_iU^\dag$, and $Z_i$ is the Pauli-$Z$ gate applied on the $i$th qubit~\cite{TM18}.
For any $1\le i\le n$, the equality $g_i|\psi_t\rangle=|\psi_t\rangle$ holds.
Furthermore, for any distinct $i$ and $j$, the operators $g_i$ and $g_j$ commute.
This is why we call these operators generalized stabilizer operators.
Unlike ordinary stabilizer operators, they cannot generally be written as a tensor product of Pauli operators.

By using the generalized stabilizer operators $\{g_i\}_{i=1}^n$, we obtain 
\begin{eqnarray*}
|\psi_t\rangle\langle\psi_t|=\prod_{i=1}^n\cfrac{I^{\otimes n}+\hat{g}_i}{2}=\cfrac{1}{2^n}\sum_{{\bf k}\in\{0,1\}^n}\hat{s}_{\bf k},
\end{eqnarray*}
where $\hat{s}_{\bf k}\equiv\prod_{i=1}^n\hat{g}_i^{k_i}$, and ${\bf k}\equiv k_1\ldots k_n\in\{0,1\}^n$.
Therefore, 
\begin{eqnarray}
\nonumber
\langle\psi_t|\hat{\rho}_{\rm out}|\psi_t\rangle&=&{\rm Tr}\left[\hat{\rho}_{\rm out}|\psi_t\rangle\langle\psi_t|\right]\\
\nonumber
&=&\cfrac{1}{2^n}\sum_{{\bf k}\in\{0,1\}^n}{\rm Tr}\left[\hat{\rho}_{\rm out}\hat{s}_{\bf k}\right]\\
\label{sampling1}
&=&\cfrac{1}{2^n}\sum_{{\bf k}\in\{0,1\}^n}{\rm Re}\left[{\rm Tr}\left[\hat{\rho}_{\rm out}\hat{s}_{\bf k}\right]\right],
\end{eqnarray}
where ${\rm Re}[c]$ is the real part of the complex number $c$, and  in the last equality, we have used the fact that $\hat{s}_{\bf k}$ is Hermitian for any ${\bf k}$.
Eq.~(\ref{sampling1}) means that we can estimate the fidelity as follows: first, a verifier generates an $n$-bit string ${\bf k}$ uniformly at random and then measures $\hat{s}_{\bf k}$ on $\hat{\rho}_{\rm out}$.
The verifier repeats this procedure polynomially many times.
A problem of this estimation method is that the measurement of $\hat{s}_{\bf k}$ generally requires an $n$-qubit operation.

\subsection{Division with respect to the space}
\label{IIIB}
Our goal is to show that Eq.~(\ref{sampling1}) can be efficiently estimated using noisy $(m+1)$-qubit and $(n-m+1)$-qubit measurements, where we assume $n/2\le m$, implying that $(n-m+1)$-qubit measurements can be implemented by using an $(m+1)$-qubit measurement device.
As the first step toward this goal, in this section, we give $(n+1)$-qubit quantum circuits to estimate Eq.~(\ref{sampling1}).

\begin{figure}[t]
\centering
\includegraphics[width=9cm, clip]{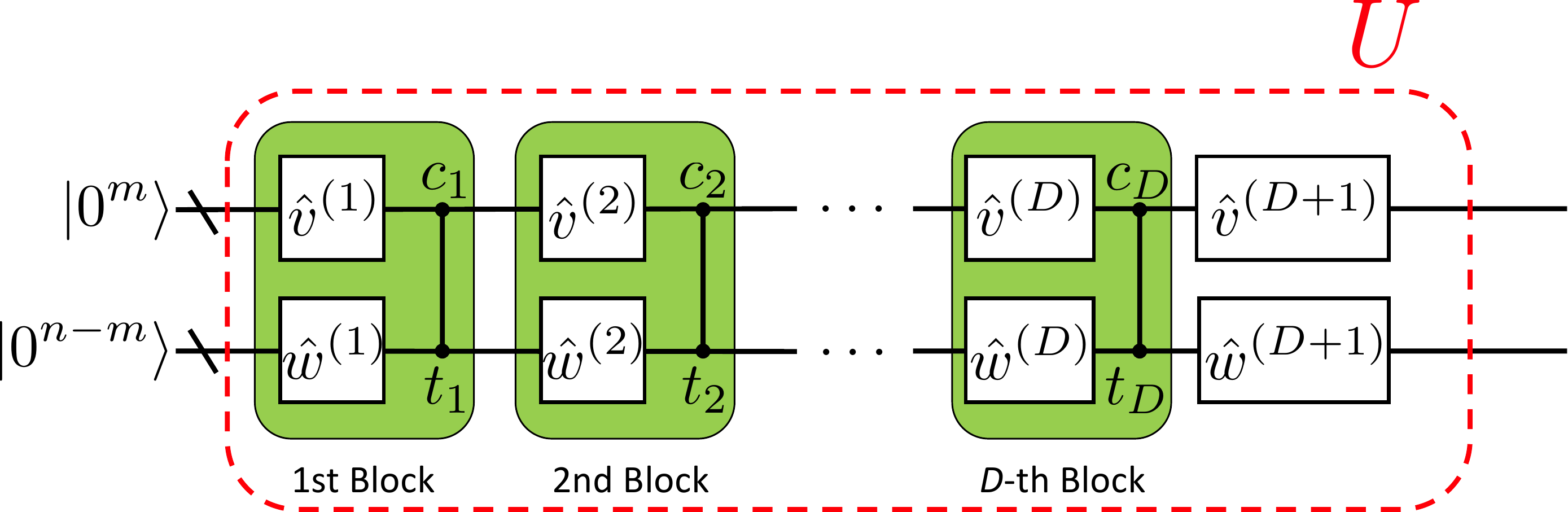}
\caption{The shallow quantum circuit $U$ on a sparse chip. Since any single-qubit gates and the $CZ$ gate constitute a universal gate set, $U$ can be decomposed as this figure. For all $1\le i\le D+1$, $\hat{v}^{(i)}$ and $\hat{w}^{(i)}$ are $m$-qubit and $(n-m)$-qubit quantum gates, respectively. Here, $D$ is the denseness of $U|0^n\rangle$. For all $1\le i\le D$, the labels $c_i$ and $t_i$ indicate on which two qubits the corresponding $CZ$ gate is applied, where $1\le c_i\le m< t_i\le n$.}
\label{decomposition}
\end{figure}

By supposing that the shallow quantum circuit $U$ on a sparse chip is composed of a  polynomial number of single-qubit gates and $CZ$ gates, it can be decomposed as shown in Fig.~\ref{decomposition}:
\begin{eqnarray*}
U&=&\left(\hat{v}^{(D+1)}\otimes\hat{w}^{(D+1)}\right){CZ}_{c_D,t_D}\left(\hat{v}^{(D)}\otimes\hat{w}^{(D)}\right)\\
&&{CZ}_{c_{D-1},t_{D-1}}\ldots\left(\hat{v}^{(1)}\otimes\hat{w}^{(1)}\right)\\
&\equiv&\left(\hat{v}^{(D+1)}\otimes\hat{w}^{(D+1)}\right)\prod_{i=D}^1{CZ}_{c_i,t_i}\left(\hat{v}^{(i)}\otimes\hat{w}^{(i)}\right),
\end{eqnarray*}
where $\hat{v}^{(i)}$ and $\hat{w}^{(i)}$ are $m$-qubit and $(n-m)$-qubit quantum gates for any $1\le i\le D+1$, respectively.
Here, $D$ is the denseness of $U|0^n\rangle$.
The quantum gate $CZ_{c_i,t_i}$ is the $CZ$ gate applied on the $c_i$th and $t_i$th qubits for any $1\le i\le D$, where $1\le c_i\le m< t_i\le n$.

To use the space-dividing technique of Bravyi {\it et al.}~\cite{BSS16}, which replaces two-qubit gates with combinations of single-qubit gates, we further decompose $CZ$ as follows:
\begin{eqnarray*}
CZ=\cfrac{1}{2}\sum_{i,j\in\{0,1\}}(-1)^{ij}\hat{\sigma}(i)\otimes\hat{\sigma}(j),
\end{eqnarray*}
where $\hat{\sigma}(0)\equiv I$ and $\hat{\sigma}(1)\equiv Z$.
Let
\begin{eqnarray}
\label{Vi}
V_{\bf i}\equiv\hat{v}^{(D+1)}\prod_{k=D}^1\hat{\sigma}_{c_k}(i_k)\hat{v}^{(k)},
\end{eqnarray}
where ${\bf i}\equiv i_1\ldots i_D\in\{0,1\}^D$.
Let
\begin{eqnarray}
\label{Wi}
W_{\bf j}\equiv\hat{w}^{(D+1)}\prod_{k=D}^1\hat{\sigma}_{t_k}(j_k)\hat{w}^{(k)}.
\end{eqnarray}
By using Eqs.~(\ref{Vi}) and (\ref{Wi}), the quantum gate $U$ is represented as
\begin{eqnarray}
\label{decom}
U=\cfrac{1}{2^D}\sum_{{\bf i},{\bf j}\in\{0,1\}^D}(-1)^{{\bf i}\cdot{\bf j}}V_{\bf i}\otimes W_{\bf j},
\end{eqnarray}
where ${\bf i}\cdot{\bf j}=\oplus_{k=1}^Di_kj_k$.

\begin{figure}[t]
\centering
\includegraphics[width=8.5cm, clip]{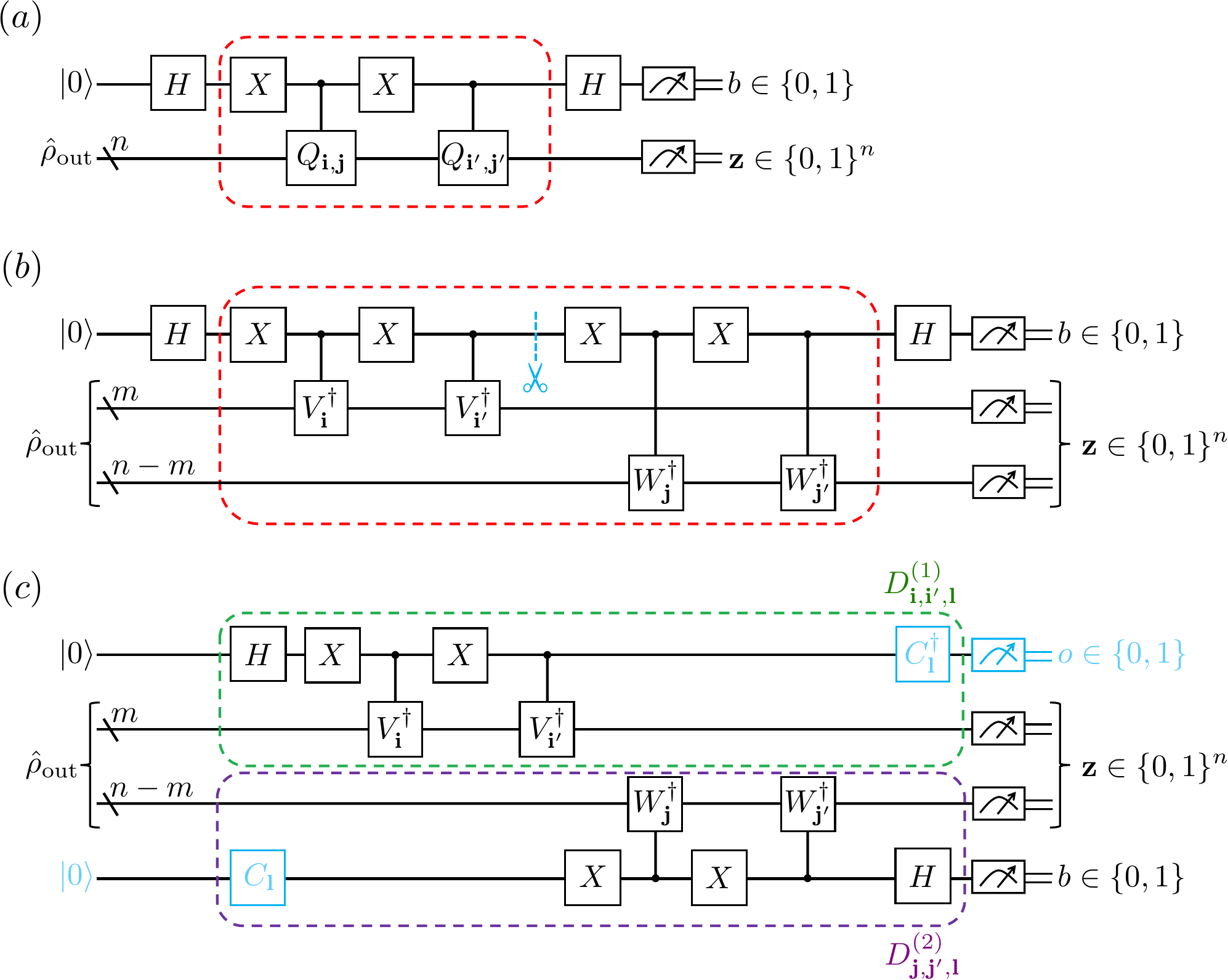}
\caption{The quantum circuits to estimate the real part of Eq.~(\ref{cross}). (a) The $(n+1)$-qubit quantum circuit used in Sec.~\ref{IIIB}. The meter symbols represent measurements in the computational basis. (b) The quantum circuit in (a) can be rewritten as this circuit. The difference is depicted by a dotted red box. Since a common ancillary qubit $|0\rangle$ is used for the first $m$ and other $(n-m)$ qubits, this circuit cannot be split into two small circuits. (c) The quantum circuit in (b) can be divided into $(m+1)$- and $(n-m+1)$-qubit quantum circuits. The dotted blue vertical line in (b) is replaced with blue operations in (c). The Clifford operations $\{C_{\bf l}\}_{{\bf l}\in\{0,1\}^3}$ are defined in Eq.~(\ref{clifford}). The operations enclosed by the green (upper) and purple (lower) boxes are denoted by $D^{(1)}_{{\bf i},{\bf i}',{\bf l}}$ and $D^{(2)}_{{\bf j},{\bf j}',{\bf l}}$, respectively.}
\label{patch}
\end{figure}

We define an $n$-qubit gate $Q_{{\bf i},{\bf j}}^\dag\equiv V_{\bf i}\otimes W_{\bf j}$.
From Eq.~(\ref{decom}),
\begin{eqnarray}
\nonumber
\hat{s}_{\bf k}&=&U\left(\prod_{i=1}^nZ_i^{k_i}\right)U^\dag\\
\label{stabi}
&=&\cfrac{1}{4^D}\sum_{{\bf i},{\bf j},{\bf i}',{\bf j}'\in\{0,1\}^D}(-1)^{{\bf i}\cdot{\bf j}+{\bf i}'\cdot{\bf j}'}\left[Q_{{\bf i},{\bf j}}^\dag\left(\prod_{i=1}^{n}Z_i^{k_i}\right)Q_{{\bf i}',{\bf j}'}\right].\ \ \ \ \
\end{eqnarray}
Therefore, ${\rm Re}[{\rm Tr}[\hat{\rho}_{\rm out}\hat{s}_{\bf k}]]$ can be estimated by estimating the real part of
\begin{eqnarray}
\label{cross}
(-1)^{{\bf i}\cdot{\bf j}+{\bf i}'\cdot{\bf j}'}{\rm Tr}\left[\hat{\rho}_{\rm out}\left[Q_{{\bf i},{\bf j}}^\dag\left(\prod_{i=1}^nZ_i^{k_i}\right)Q_{{\bf i}',{\bf j}'}\right]\right].
\end{eqnarray}
By using the quantum circuit in Fig.~\ref{patch} (a), we can estimate the real part of Eq.~(\ref{cross}).
Let $b\in\{0,1\}$ and ${\bf z}\in\{0,1\}^n$ be outputs of the quantum circuit in Fig.~\ref{patch} (a).
$\alpha({\bf i},{\bf j},{\bf i}',{\bf j}')\in\{1,-1\}$ is a random variable such that $\alpha({\bf i},{\bf j},{\bf i}',{\bf j}')=1$ if and only if $\oplus_{i=1}^nz_ik_i={\bf i}\cdot{\bf j}\oplus{\bf i}'\cdot{\bf j}'\oplus b$.
When we repeatedly run the quantum circuit in Fig.~\ref{patch} (a), the mean value of $\alpha({\bf i},{\bf j},{\bf i}',{\bf j}')$ converges to the real part of Eq.~(\ref{cross}) as shown in Ref.~\cite{BSS16}.
For the completeness of our paper, we show it in Appendix~\ref{A}.

\subsection{Division with respect to the time}
\label{IIIC}
The quantum circuit in Fig.~\ref{patch} (a) cannot be split into two small circuits because a common ancillary qubit $|0\rangle$ is used for the first $m$ and other $(n-m)$ qubits (see Fig.~\ref{patch} (b)).
To solve this problem, we utilize the time-dividing technique of Peng {\it et al.}~\cite{PHOW19}, which replaces the identity gate with a combination of single-qubit measurements and preparations.

We replace the identity gate (the dotted blue vertical line in Fig.~\ref{patch} (b)) with Pauli-basis measurements $\{C_{\bf l}|0\rangle\langle 0|C_{\bf l}^\dag,C_{\bf l}|1\rangle\langle 1|C_{\bf l}^\dag\}_{{\bf l}\in\{0,1\}^3}$ and the preparation of Pauli-basis states $\{C_{\bf l}|0\rangle\}_{{\bf l}\in\{0,1\}^3}$ (blue parts in Fig.~\ref{patch} (c)).
Here, 
\begin{eqnarray}
\label{clifford}
C_{\bf l}\equiv S^{\delta_{l_1,1}\delta_{l_2,0}}H^{\delta_{l_1+l_2,1}}X^{l_3}
\end{eqnarray}
is the Clifford operation, where $S\equiv\sqrt{Z}$, $H$ is the Hadamard gate, $X$ is the Pauli-$X$ gate, and $\delta_{i,j}=1$ if and only if $i=j$.
Due to this replacement, we can divide the quantum circuit in Fig.~\ref{patch} (b) into $(m+1)$- and $(n-m+1)$-qubit quantum circuits as shown in Fig.~\ref{patch} (c).

\begin{algorithm}[H]
\begin{flushleft}
\caption{Divide-and-conquer verification method}
{\bf Purpose:} For desired $0<\epsilon,\delta<1$, and any $n$-qubit sparse state $|\psi_t\rangle=U|0^n\rangle$, output $F_{\rm est}$ such that with probability of at least $1-\delta$, $|F_{\rm est}-\langle\psi_t|\hat{\rho}_{\rm out}|\psi_t\rangle|\le\epsilon$.\\
\noindent
{\bf Assumption:} The actual output state $\hat{\rho}_{\rm out}$ is identically and independently (i.i.d.) distributed, i.e., the actual quantum circuit always outputs $\hat{\rho}_{\rm out}$.
Any $(m+1)$-qubit unitary gate can be approximately achieved in the sense that the diamond norm~\cite{AKN97} between its ideal operation and its actual one is at most $\epsilon/4^{D+2}$, where $n/2\le m<n-1$ and $D$ is explained later.\\
\noindent
{\bf Procedures:}
\begin{enumerate}
\item For a target quantum gate $U$ that is supposed to be implemented by the ideal quantum circuit, divide $n$ input qubits into $m$ and $(n-m)$ qubits such that the number of $CZ$ gates between them is $D=O(\log{n})$.
This division is possible because of the sparsity of $|\psi_t\rangle$.
\item Under the division obtained in step 1, decompose $U$ as Eq.~(\ref{decom}).
\item Choose ${\bf k}\in\{0,1\}^{n}$ $T_1$ times uniformly at random, and then perform the following steps for each ${\bf k}$.
\begin{enumerate}
\item Calculate the generalized stabilizer operator $\hat{s}_{\bf k}$ (see Eq.~(\ref{stabi})).
\item Choose quadruples $({\bf i},{\bf j},{\bf i}',{\bf j}')$ $T_2$ times uniformly at random. Then, perform the following step for each quadruple $({\bf i},{\bf j},{\bf i}',{\bf j}')$.
\begin{enumerate}
\item[({\rm$\alpha$})] Estimate the expected value of $\sum_{{\bf l}\in\{0,1\}^3}\beta({\bf i},{\bf j},{\bf i}',{\bf j}',{\bf l})/2$ as an estimated value of the real part of Eq.~(\ref{cross}).
To this end, run the quantum circuit in Fig.~\ref{patch} (c) $T_3$ times for each ${\bf l}\in\{0,1\}^3$.
\end{enumerate}
\item By averaging the $T_2$ estimated values of the real part of Eq.~(\ref{cross}) obtained in step ({\rm$\alpha$}), derive an estimated value of ${\rm Tr}[\hat{\rho}_{\rm out}\hat{s}_{\bf k}]$.
\end{enumerate}
\item By averaging the $T_1$ estimated values of ${\rm Tr}[\hat{\rho}_{\rm out}\hat{s}_{\bf k}]$, obtain $F_{\rm est}$.
\end{enumerate}
\end{flushleft}
\end{algorithm}

We define a random variable $\beta({\bf i},{\bf j},{\bf i}',{\bf j}',{\bf l})\equiv(-1)^{(1+\delta_{l_1,0}\delta_{l_2,0})o}\alpha({\bf i},{\bf j},{\bf i}',{\bf j}')$, where $o$ is the outcome of the first qubit in Fig.~\ref{patch} (c).
The sum $\sum_{{\bf l}\in\{0,1\}^3}\beta({\bf i},{\bf j},{\bf i}',{\bf j}',{\bf l})/2$ converges to the real part of Eq.~(\ref{cross}) as shown in Ref.~\cite{PHOW19}.
For the completeness of our paper, we show it in Appendix~\ref{B}.

The procedures of our verification method are summarized as Algorithm~1.
In Algorithm~1, the values of $T_1$, $T_2$, and $T_3$ remain free, which determine the sample complexity of our method.
These values depend on the accuracy $\epsilon$ and the success probability $1-\delta$ and will be derived in the next subsection.

\subsection{Sample complexity}
\label{IIID}
In this section, we derive the sample complexity of our method under the assumption that any $(m+1)$-qubit gate can be achieved with the diamond distance at most $\epsilon/4^{D+2}$, and $T_1=T_2=T_3$.
The sample complexity is the number of copies of $\hat{\rho}_{\rm out}$ required to obtain $F_{\rm est}$ such that with probability of at least $1-\delta$, $|F_{\rm est}-\langle\psi_t|\hat{\rho}_{\rm out}|\psi_t\rangle|\le\epsilon$.
More precisely, we show the following theorem:
\begin{theorem}
\label{theorem}
Let $|\psi_t\rangle\equiv U|0^n\rangle$ be an $n$-qubit sparse state, where $U$ consists of a polynomial number of $CZ$ gates and single-qubit quantum gates.
Suppose that $n$ qubits are divided into $m$ and $(n-m)$ qubits, where $n/2\le m$, $m=\Theta(n)$, and $n-m=\Theta(n)$, such that the number of $CZ$ gates between them is $D=O(\log{n})$.
We assume that $T_1=T_2=T_3=T$, and the diamond norm between any ideal and actual $(m+1)$-qubit gates is upper bounded by $\epsilon/4^{D+2}$.
Then for any $n$-qubit state $\hat{\rho}_{\rm out}$, Algorithm~1 outputs $F_{\rm est}$ such that with probability of at least $1-\delta$, $|F_{\rm est}-\langle\psi_t|\hat{\rho}_{\rm out}|\psi_t\rangle|\le \epsilon$, by performing $(m+1)$-qubit measurements on
\begin{eqnarray*}
8T^3=O\left(\cfrac{2^{12D}}{\epsilon^6}\left(D+\log{\cfrac{1}{\delta\epsilon^4}}\right)^3\right)
\end{eqnarray*}
copies of $\hat{\rho}_{\rm out}$, where $0<\epsilon,\delta<1$.
\end{theorem}
{\it Proof.} Our verification method consist of three parts: steps 1-4, steps (a)-(c), and step ({\rm $\alpha$}).

First, we consider step ({\rm $\alpha$}).
The ideal functionalities of green and purple dotted boxes in Fig.~\ref{patch} (c) are represented by $D^{(1)}_{{\bf i},{\bf i}',{\bf l}}$ and $D^{(2)}_{{\bf j},{\bf j}',{\bf l}}$, respectively.
The actual functionalities somewhat deviate from the ideal ones.
We represent the actual functionalities by completely positive trace-preserving (CPTP) maps $\mathcal{E}^{(1)}_{{\bf i},{\bf i}',{\bf l}}$ and $\mathcal{E}^{(2)}_{{\bf j},{\bf j}',{\bf l}}$, respectively.
Without loss of generality, we can assume that $Z$-basis measurements and the initialization to $|0\rangle$ in Fig.~\ref{patch} (c) can be perfectly implemented, because their imperfections can be treated as imperfections on quantum gates $D^{(1)}_{{\bf i},{\bf i}',{\bf l}}$ and $D^{(2)}_{{\bf j},{\bf j}',{\bf l}}$.
For convenience, we define a super-operator $\mathcal{U}(\cdot)\equiv U(\cdot)U^\dag$ for any unitary operator $U$.
From the assumption in Algorithm~1, there exists $\epsilon_{\diamond}(\le\epsilon/4^{D+2})$ such that
\begin{eqnarray*}
\left|\left|\mathcal{D}^{(1)}_{{\bf i},{\bf i}',{\bf l}}-\mathcal{E}^{(1)}_{{\bf i},{\bf i}',{\bf l}}\right|\right|_{\diamond}&\le&\epsilon_{\diamond}\\
\left|\left|\mathcal{D}^{(2)}_{{\bf j},{\bf j}',{\bf l}}-\mathcal{E}^{(2)}_{{\bf j},{\bf j}',{\bf l}}\right|\right|_{\diamond}&\le&\epsilon_{\diamond}
\end{eqnarray*}
for all ${\bf i}$, ${\bf i}'$, ${\bf j}$, ${\bf j}'$, and ${\bf l}$.

Let $\Pi_{\beta=1}$ and $\Pi_{\beta=-1}$ be projectors onto the space satisfying $\beta({\bf i},{\bf j},{\bf i}',{\bf j}',{\bf l})=1$ and $\beta({\bf i},{\bf j},{\bf i}',{\bf j}',{\bf l})=-1$, respectively.
We define an observable $B\equiv\Pi_{\beta=1}-\Pi_{\beta=-1}$.
We also define $\hat{\rho}'_{\rm out}\equiv|0\rangle\langle0|\otimes\hat{\rho}_{\rm out}\otimes|0\rangle\langle0|$.
By using them, we show that the expected value of $B$ obtained in the actual case is close to that in the ideal case.
More formally, we obtain
\begin{eqnarray}
\nonumber
&&\left|{\rm Tr}\left[\left(\mathcal{D}^{(1)}_{{\bf i},{\bf i}',{\bf l}}\otimes\mathcal{D}^{(2)}_{{\bf j},{\bf j}',{\bf l}}-\mathcal{E}^{(1)}_{{\bf i},{\bf i}',{\bf l}}\otimes\mathcal{E}^{(2)}_{{\bf j},{\bf j}',{\bf l}}\right)(\hat{\rho}'_{\rm out})B\right]\right|\\
\nonumber
&\le&\sum_{a\in\{1,-1\}}\\
\nonumber
&&\left|{\rm Tr}\left[\left(\mathcal{D}^{(1)}_{{\bf i},{\bf i}',{\bf l}}\otimes\mathcal{D}^{(2)}_{{\bf j},{\bf j}',{\bf l}}-\mathcal{E}^{(1)}_{{\bf i},{\bf i}',{\bf l}}\otimes\mathcal{E}^{(2)}_{{\bf j},{\bf j}',{\bf l}}\right)(\hat{\rho}'_{\rm out})\Pi_{\beta=a}\right]\right|\\
\nonumber
&\le&\left|\left|\left(\mathcal{D}^{(1)}_{{\bf i},{\bf i}',{\bf l}}\otimes\mathcal{D}^{(2)}_{{\bf j},{\bf j}',{\bf l}}-\mathcal{E}^{(1)}_{{\bf i},{\bf i}',{\bf l}}\otimes\mathcal{E}^{(2)}_{{\bf j},{\bf j}',{\bf l}}\right)(\hat{\rho}'_{\rm out})\right|\right|_1\\
\nonumber
&\le&\left|\left|\mathcal{D}^{(1)}_{{\bf i},{\bf i}',{\bf l}}\otimes\mathcal{D}^{(2)}_{{\bf j},{\bf j}',{\bf l}}-\mathcal{E}^{(1)}_{{\bf i},{\bf i}',{\bf l}}\otimes\mathcal{E}^{(2)}_{{\bf j},{\bf j}',{\bf l}}\right|\right|_\diamond\\
\nonumber
&\le&\left|\left|\left(\mathcal{E}^{(1)}_{{\bf i},{\bf i}',{\bf l}}-\mathcal{D}^{(1)}_{{\bf i},{\bf i}',{\bf l}}\right)\otimes\mathcal{E}^{(2)}_{{\bf j},{\bf j}',{\bf l}}\right|\right|_\diamond\\
\nonumber
&&+\left|\left|\mathcal{D}^{(1)}_{{\bf i},{\bf i}',{\bf l}}\otimes\left(\mathcal{E}^{(2)}_{{\bf j},{\bf j}',{\bf l}}-\mathcal{D}^{(2)}_{{\bf j},{\bf j}',{\bf l}}\right)\right|\right|_\diamond\\
\nonumber
&=&\left|\left|\mathcal{E}^{(1)}_{{\bf i},{\bf i}',{\bf l}}-\mathcal{D}^{(1)}_{{\bf i},{\bf i}',{\bf l}}\right|\right|_\diamond\left|\left|\mathcal{E}^{(2)}_{{\bf j},{\bf j}',{\bf l}}\right|\right|_\diamond\\
\nonumber
&&+\left|\left|\mathcal{D}^{(1)}_{{\bf i},{\bf i}',{\bf l}}\right|\right|_\diamond\left|\left|\mathcal{E}^{(2)}_{{\bf j},{\bf j}',{\bf l}}-\mathcal{D}^{(2)}_{{\bf j},{\bf j}',{\bf l}}\right|\right|_\diamond\\
\label{diamond}
&\le&2\epsilon_{\diamond},
\end{eqnarray}
where $||\cdot||_1$ and $||\cdot||_\diamond$ are the Schatten $1$-norm and the diamond norm, respectively.
We have used the triangle inequality to derive the first and fourth inequalities.
The second and third inequalities come from Theorem 9.1 in Ref.~\cite{NC00} and the definition of the diamond norm, respectively.
The fifth equality is the multiplicativity of the diamond norm with respect to tensor products~\cite{F09}.

A verifier repeatedly runs the quantum circuit in Fig.~\ref{patch} (c) $T$ times for each of chosen ${\bf i},{\bf i}',{\bf j},{\bf j}',{\bf l}$.
Note that the verifier can implement only an approximated quantum circuit, i.e., the circuit in Fig.~\ref{patch} (c) with $\mathcal{E}^{(1)}_{{\bf i},{\bf i}',{\bf l}}$ and $\mathcal{E}^{(2)}_{{\bf j},{\bf j}',{\bf l}}$.
Let $\beta'_k({\bf i},{\bf j},{\bf i}',{\bf j}',{\bf l})$ be the $k$th random variable obtained from such approximated quantum circuit for $1\le k\le T$.
From the Hoeffding inequality~\cite{H63},
\begin{eqnarray*}
&&\left|\cfrac{\sum_{k=1}^T\beta'_k({\bf i},{\bf j},{\bf i}',{\bf j}',{\bf l})}{T}-{\rm Tr}\left[\mathcal{E}^{(1)}_{{\bf i},{\bf i}',{\bf l}}\otimes\mathcal{E}^{(2)}_{{\bf j},{\bf j}',{\bf l}}(\hat{\rho}'_{\rm out})B\right]\right|\\
&\le&\epsilon_h
\end{eqnarray*}
holds with probability of at least $1-2e^{-T\epsilon_h^2/2}$, where $\epsilon_h>0$.
Therefore, from Eq.~(\ref{diamond}), with probability of at least $1-2e^{-T\epsilon_h^2/2}$,
\begin{eqnarray}
\nonumber
&&\left|\cfrac{\sum_{k=1}^T\beta'_k({\bf i},{\bf j},{\bf i}',{\bf j}',{\bf l})}{T}-{\rm Tr}\left[\mathcal{D}^{(1)}_{{\bf i},{\bf i}',{\bf l}}\otimes\mathcal{D}^{(2)}_{{\bf j},{\bf j}',{\bf l}}(\hat{\rho}'_{\rm out})B\right]\right|\\
\label{bound1}
&\le&\epsilon_h+2\epsilon_{\diamond}.
\end{eqnarray}
Since $\sum_{{\bf l}\in\{0,1\}^3}\beta({\bf i},{\bf j},{\bf i}',{\bf j}',{\bf l})/2$ obtained from the ideal circuit in Fig.~\ref{patch} (c) converges to the real part of Eq.~(\ref{cross}) as shown in Appendix~\ref{B}, from Eq.~(\ref{bound1}),
\begin{widetext}
\begin{eqnarray}
\nonumber
&&\left|\sum_{{\bf l}\in\{0,1\}^3}\cfrac{\sum_{k=1}^T\beta'_k({\bf i},{\bf j},{\bf i}',{\bf j}',{\bf l})}{2T}-{\rm Re}\left[(-1)^{{\bf i}\cdot{\bf j}+{\bf i}'\cdot{\bf j}'}{\rm Tr}\left[\hat{\rho}_{\rm out}\left[Q_{{\bf i},{\bf j}}^\dag\left(\prod_{i=1}^nZ_i^{k_i}\right)Q_{{\bf i}',{\bf j}'}\right]\right]\right]\right|\\
\nonumber
&=&\left|\sum_{{\bf l}\in\{0,1\}^3}\cfrac{\sum_{k=1}^T\beta'_k({\bf i},{\bf j},{\bf i}',{\bf j}',{\bf l})}{2T}-\cfrac{1}{2}\sum_{{\bf l}\in\{0,1\}^3}{\rm Tr}\left[\mathcal{D}^{(1)}_{{\bf i},{\bf i}',{\bf l}}\otimes\mathcal{D}^{(2)}_{{\bf j},{\bf j}',{\bf l}}(\hat{\rho}'_{\rm out})B\right]\right|\\
\nonumber
&\le&\cfrac{1}{2}\sum_{{\bf l}\in\{0,1\}^3}\Bigg|\cfrac{\sum_{k=1}^T\beta'_k({\bf i},{\bf j},{\bf i}',{\bf j}',{\bf l})}{T}-{\rm Tr}\left[\mathcal{D}^{(1)}_{{\bf i},{\bf i}',{\bf l}}\otimes\mathcal{D}^{(2)}_{{\bf j},{\bf j}',{\bf l}}(\hat{\rho}'_{\rm out})B\right]\Bigg|\\
\label{bound2}
&\le&4\epsilon_h+8\epsilon_{\diamond}
\end{eqnarray}
\end{widetext}
with probability of at least $(1-2e^{-T\epsilon_h^2/2})^8$.

Second, we consider steps (a)-(c).
In these steps, the verifier chooses quadruples $({\bf i},{\bf j},{\bf i}',{\bf j}')$ $T$ times.
Let $({\bf i}^{(i)},{\bf j}^{(i)},{\bf i}^{'(i)},{\bf j}^{'(i)})$ be the $i$th quadruple for $1\le i\le T$.
For simplicity, we define
\begin{eqnarray*}
&&q({\bf i},{\bf j},{\bf i}',{\bf j}',{\bf k})\\
&\equiv&{\rm Re}\left[(-1)^{{\bf i}\cdot{\bf j}+{\bf i}'\cdot{\bf j}'}{\rm Tr}\left[\hat{\rho}_{\rm out}\left[Q_{{\bf i},{\bf j}}^\dag\left(\prod_{i=1}^nZ_i^{k_i}\right)Q_{{\bf i}',{\bf j}'}\right]\right]\right].
\end{eqnarray*}
From the Hoeffding inequality and Eq.~(\ref{stabi}),
\begin{eqnarray}
\nonumber
\left|\cfrac{1}{T}\sum_{i=1}^Tq({\bf i}^{(i)},{\bf j}^{(i)},{\bf i}^{'(i)},{\bf j}^{'(i)},{\bf k})-\cfrac{{\rm Tr}[\hat{\rho}_{\rm out}\hat{s}_{\bf k}]}{4^D}\right|\le\epsilon_h\\
\label{bound3}
\end{eqnarray}
with probability of at least $1-2e^{-T{\epsilon_h}^2/2}$.

Let $\tilde{\beta}({\bf i},{\bf i}',{\bf j},{\bf j}')\equiv\sum_{{\bf l}\in\{0,1\}^3}\left(\sum_{k=1}^T\beta'_k({\bf i},{\bf j},{\bf i}',{\bf j}',{\bf l})\right)/(2T)$.
By using the triangle inequality with Eqs.~(\ref{bound2}) and (\ref{bound3}),
\begin{eqnarray*}
&&\left|4^D\cfrac{\sum_{i=1}^T\tilde{\beta}({\bf i}^{(i)},{\bf i}^{'(i)},{\bf j}^{(i)},{\bf j}^{'(i)})}{T}-{\rm Tr}[\hat{\rho}_{\rm out}\hat{s}_{\bf k}]\right|\\
&\le&4^D\left(5\epsilon_h+8\epsilon_\diamond\right)
\end{eqnarray*}
with probability of at least $(1-2e^{-T{\epsilon_h}^2/2})^{8T+1}$.

Finally, we consider steps 1-4.
In these steps, the verifier chooses the bit string ${\bf k}$ $T$ times.
Let ${\bf k}^{(i)}$ be the $i$th chosen bit string for $1\le i\le T$.
From the Hoeffding inequality and Eq.~(\ref{sampling1}),
\begin{eqnarray*}
&&{\rm Pr}\left[\Bigg|\cfrac{\sum_{i=1}^T{\rm Tr}\left[\hat{\rho}_{\rm out}\hat{s}_{{\bf k}^{(i)}}\right]}{T}-\langle\psi_t|\hat{\rho}_{\rm out}|\psi_t\rangle\Bigg|\le\epsilon_h\right]\\
&\ge&1-2e^{-T{\epsilon_h}^2/2}.
\end{eqnarray*}
Let $f\equiv4^D\left(\sum_{i=1}^T\tilde{\beta}({\bf i}^{(i)},{\bf i}^{'(i)},{\bf j}^{(i)},{\bf j}^{'(i)})\right)/T$.
We also define $f^{(i)}$ as the random variable $f$ obtained in the case of ${\bf k}^{(i)}$.
Therefore, by using the triangle inequality, with the probability of at least $(1-2e^{-T{\epsilon_h}^2/2})^{8T^2+T+1}$,
\begin{eqnarray*}
\left|\cfrac{\sum_{i=1}^{T}f^{(i)}}{T}-\langle\psi_t|\hat{\rho}_{\rm out}|\psi_t\rangle\right|\le4^D\left(5\epsilon_h+8\epsilon_\diamond\right)+\epsilon_h.
\end{eqnarray*}
By setting $F_{\rm est}=(\sum_{i=1}^Tf^{(i)})/T$, we can estimate the value of the fidelity $\langle\psi_t|\hat{\rho}_{\rm out}|\psi_t\rangle$.

The remaining task is to calculate the value of $8T^3$ under two conditions
\begin{eqnarray}
\label{condition1}
4^D\left(5\epsilon_h+8\epsilon_\diamond\right)+\epsilon_h&\le&\epsilon\\
\label{condition2}
(1-2e^{-T{\epsilon_h}^2/2})^{8T^2+T+1}&\ge&1-\delta.
\end{eqnarray}
From $\epsilon_\diamond\le\epsilon/4^{D+2}$, we can set $\epsilon_h=\epsilon/[2(5\cdot4^D+1)]$ to satisfy Eq.~(\ref{condition1}).
Therefore, we have
\begin{eqnarray*}
&&(1-2e^{-T{\epsilon_h}^2/2})^{8T^2+T+1}\\
&\ge&1-2(8T^2+T+1)e^{-T{\epsilon_h}^2/2}\\
&\ge&1-20T^2e^{-T{\epsilon_h}^2/2}\\
&\ge&1-\cfrac{640}{{\epsilon_h}^4}e^{-T{\epsilon_h}^2/4},
\end{eqnarray*}
where we have used $T\ge1$ and $T^2\le32e^{T{\epsilon_h}^2/4}/{\epsilon_h}^4$ to derive the second and third inequalities, respectively.
To satisfy Eq.~(\ref{condition2}), it is sufficient to set $\delta=640/({\epsilon_h}^4e^{T{\epsilon_h}^2/4})$.
As a result,
\begin{eqnarray*}
T&=&\cfrac{4}{{\epsilon_h}^2}\log{\cfrac{640}{\delta{\epsilon_h}^4}}\\
&=&\cfrac{16(5\cdot4^D+1)^2}{\epsilon^2}\log{\cfrac{10240(5\cdot4^D+1)^4}{\delta\epsilon^4}}\\
&=&O\left(\cfrac{2^{4D}}{\epsilon^2}\left(D+\log{\cfrac{1}{\delta\epsilon^4}}\right)\right).
\end{eqnarray*}
This completes the proof of Theorem~\ref{theorem}.
\hspace{\fill}$\blacksquare$

\section{Proof-of-principle experiment}
\label{addIV}
In the previous section, we have assumed $\epsilon_\diamond\le\epsilon/4^{D+2}$.
It implies that high-precision quantum gates are necessary to construct $\mathcal{D}^{(1)}_{{\bf i},{\bf i}',{\bf l}}$ and $\mathcal{D}^{(2)}_{{\bf j},{\bf j}',{\bf l}}$.
As a concrete example, suppose that the quantum chip is linear, the required precision $\epsilon$ is $0.1$, and the target state $|\psi_t\rangle$ is the $n$-qubit linear graph state $\prod_{i=1}^{n-1}CZ_{i,i+1}(|+\rangle^{\otimes n})$ with $n$ being even.
Here, $|+\rangle\equiv(|0\rangle+|1\rangle)/\sqrt{2}$.
In this case, each of $\mathcal{D}^{(1)}_{{\bf i},{\bf i}',{\bf l}}$ and $\mathcal{D}^{(2)}_{{\bf j},{\bf j}',{\bf l}}$ can be constructed by using $(13n-15)$ single-qubit gates and $(8n-10)$ $CZ$'s.
Therefore, the diamond distance between ideal and actual elementary gates must be at most $1/[640(21n-25)]$.
Even when $n=2$, this value $\sim0.009\%$ is extremely low.
Note that for simplicity, we assume that $Z$-basis measurements and the initialization to $|0\rangle$ are perfectly implemented.

\begin{figure}[t]
\centering
\includegraphics[width=9cm, clip]{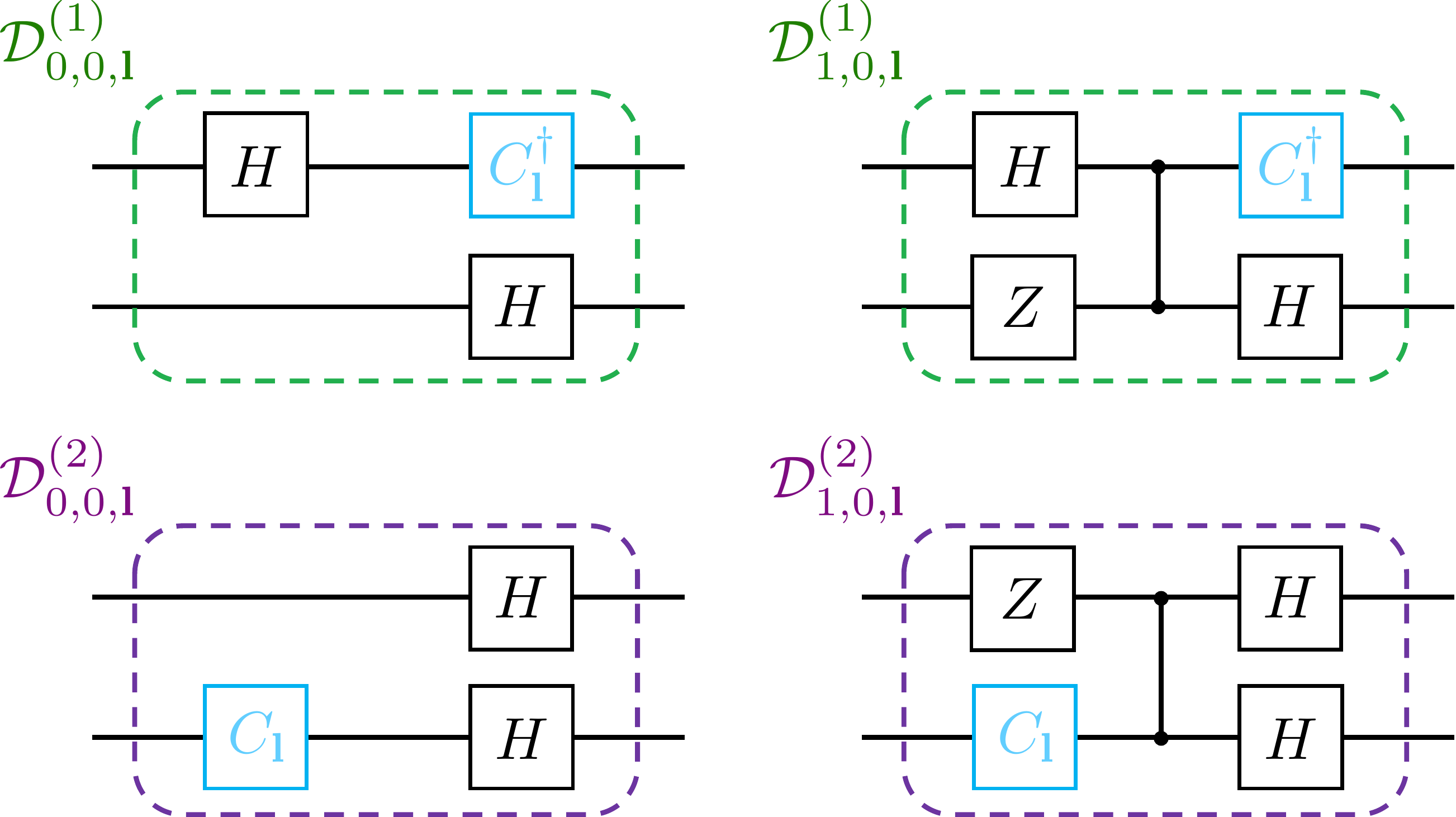}
\caption{Circuit diagrams of $\mathcal{D}^{(1)}_{i,0,{\bf l}}$ and $\mathcal{D}^{(2)}_{j,0,{\bf l}}$ for $i,j\in\{0,1\}$ in the case of $|\psi_t\rangle=CZ|+\rangle^{\otimes 2}$. Here, $H$ is the Hadamard gate. These circuits are obtained by simplifying circuits in Fig.~\ref{patch} (c). To this end, we use two equalities $(I\otimes X)CZ(Z\otimes X)=CZ$ and $CH(X\otimes I)CH(X\otimes I)=I\otimes H$, where $CH\equiv|0\rangle\langle 0|\otimes I+|1\rangle\langle 1|\otimes H$.}
\label{simplified}
\end{figure}

However, the above analysis is too pessimistic in several situations.
Our analysis does not take an actual error model into account (e.g., see calculations to derive Eq.~(\ref{diamond})).
Furthermore, once $|\psi_t\rangle$ is fixed, we may be able to use properties of $|\psi_t\rangle$ to simplify quantum circuits in Fig.~\ref{patch} (c).
To observe the actual performance of our method, we perform a $4$-qubit experiment on the IBM Manila 5-qubit chip.
The target state is $|\psi_t\rangle=CZ|+\rangle^{\otimes 2}$.
In this case, the fidelity $\langle\psi_t|\hat{\rho}_{\rm out}|\psi_t\rangle$ can be written as
\begin{eqnarray}
\nonumber
\cfrac{1}{4}\sum_{i,j\in\{0,1\}}(-1)^{ij}{\rm Re}\left[{\rm Tr}\left[\hat{\rho}_{\rm out}\left[Q_{i,j}^\dag\left(Z_1^j\otimes Z_2^i\right)Q_{0,0}\right]\right]\right],\\
\label{ex1}
\end{eqnarray}
where we have used
\begin{eqnarray*}
&&Q_{i,j}^\dag\left(Z_1^{k_1}\otimes Z_2^{k_2}\right)Q_{i',j'}\\
&=&(-1)^{i'k_1+j'k_2}Q_{i\oplus i',j\oplus j'}^\dag\left(Z_1^{k_1}\otimes Z_2^{k_2}\right)Q_{0,0}
\end{eqnarray*}
for any $i,j,i',j'\in\{0,1\}$ and ${\bf k}\in\{0,1\}^2$.
In our experiment, we estimate each term by running quantum circuits in Fig.~\ref{patch} (c) $T=1024$ times for each ${\bf l}\in\{0,1\}^3$.
In other words, we calculate
\begin{eqnarray*}
F_{\rm est}=\cfrac{1}{4}\sum_{i,j\in\{0,1\}}\sum_{{\bf l}\in\{0,1\}^3}\cfrac{\sum_{k=1}^{1024}\beta'_k(i,j,0,0,{\bf l})}{2048}
\end{eqnarray*}
as an estimated value of Eq.~(\ref{ex1}).
To this end, we simplify $\mathcal{D}^{(1)}_{i,0,{\bf l}}$ and $\mathcal{D}^{(2)}_{j,0,{\bf l}}$ as given in Fig.~\ref{simplified}.
Depending on ${\bf l}$, we further simplify the circuits in Fig.~\ref{simplified}.
For example, when ${\bf l}=000$, the $CZ$ gate in $\mathcal{D}_{1,0,{\bf l}}^{(2)}$ can be removed because $C_{000}|0\rangle=|0\rangle$.
By running these simplified circuits on the IBM Manila 5-qubit chip, we obtain $F_{\rm est}\simeq0.789$.
The experimental data is given in Appendix~\ref{AC}.

To check the accuracy of $F_{\rm est}$, a true value $F\equiv\langle\psi_t|\hat{\rho}_{\rm out}|\psi_t\rangle$ is necessary.
However, in principle, it should be hard to derive $F$.
Therefore, we alternatively compare $F_{\rm est}$ with another estimated value $\tilde{F}_{\rm est}$ that is obtained by directly estimating ${\rm Re}[{\rm Tr}[\hat{\rho}_{\rm out}\hat{s}_{\bf k}]]$ for each ${\bf k}\in\{0,1\}^2$ (see Eq.~(\ref{sampling1})).
We expect that $\tilde{F}_{\rm est}$ tends to be closer to $F$ than $F_{\rm est}$ is because quantum circuits used to obtain $\tilde{F}_{\rm est}$ are simpler than those used to obtain $F_{\rm est}$ (see Appendix~\ref{AC}).
In the case of $|\psi_t\rangle=CZ|+\rangle^{\otimes 2}$, we obtain $\hat{s}_{00}=I^{\otimes 2}$, $\hat{s}_{01}=Z\otimes X$, $\hat{s}_{10}=X\otimes Z$, and $\hat{s}_{11}=Y\otimes Y$, where $Y$ is the Pauli-$Y$ operator.
Note that since $\hat{s}_{00}=I^{\otimes 2}$, we can set ${\rm Re}[{\rm Tr}[\hat{\rho}_{\rm out}\hat{s}_{00}]]=1$ without performing experiments.
By measuring the observable $\hat{s}_{\bf k}$ $1024$ times for each ${\bf k}\in\{01,10,11\}$, we obtain $\tilde{F}_{\rm est}\simeq0.865$ (see also Appendix~\ref{AC}).
Contrary to the expectation from the above pessimistic analysis, we conclude that our method faithfully works when $n=2$ because $|F_{\rm est}-\tilde{F}_{\rm est}|<0.076$ is sufficiently small (compared with $\epsilon=0.1$).
It would be interesting to explore in which cases our method becomes particularly practical.
We leave it as a future work.

\section{Conclusion and Discussion}
\label{IV}
We have proposed and experimentally performed an efficient divide-and-conquer method to verify noisy intermediate-scale quantum (NISQ) computers.
Although we have simply divided a verification circuit into two small circuits (see Fig.~\ref{patch} (c)), our idea can be straightforwardly generalized so that a verification circuit is divided into $M$ circuits for any natural number $M$.
In this case, the sample complexity grows at least exponentially with $\Theta(\tilde{D})$, where $\tilde{D}$ is the number of $CZ$ gates across the $M$ small circuits.
Therefore, to make the sample complexity some polynomial in $n$, $M=O(\log{n})$ is required because $M-1\le\tilde{D}$.

For simplicity, we have assumed that $\hat{\rho}_{\rm out}$ is i.i.d. distributed.
This assumption may be removed by using the quantum de Finetti theorem for the fully one-way local operations and classical communication (LOCC) norm~\cite{LS15} as in previous studies~\cite{MTH17,TM18}.
Its rigorous analysis is left for future work.

We have also assumed that the quantum computing chip $\mathcal{C}$ is sparse.
Our method is superior to the direct fidelity estimation method~\cite{FL11} even when $\mathcal{C}$ is not sparse but is a planar graph with a constant maximum degree.
Since the graphs underlying almost all current physical chips are planar ones with constant maximum degrees, this assumption is quite natural.
For example, the geometry of Google's Sycamore 53-qubit chip~\cite{google} is a planar graph with the maximum degree four, although it is not sparse.
As a consequence of the planar separator theorem~\cite{LT79,LT80}, $D$ is $O(\sqrt{n}\log{n})$ for logarithmic-depth quantum circuits on any planar graph with a constant maximum degree.
Therefore, the sample complexity of our method is $2^{O(\sqrt{n}\log{n})}$, which is less than $O(2^n)$.

The core of our method is to divide a large quantum circuit into two small quantum circuits.
By performing classical post-processing on their measurement outcomes, we estimate the value of the fidelity.
Such a divide-and-conquer technique has already been used to embed a large problem into NISQ computers~\cite{FMMN20,TTLSM20} and reduce the number of two-qubit gates in quantum circuits~\cite{MF19,MF20}.
Furthermore, the technique was improved by Perlin {\it et al.}~\cite{PSSO20} and experimentally demonstrated by Ayral {\it et al.}~\cite{ARSAS20}.
Our results show that a similar technique can also be used to design an efficient verification method for NISQ computers.
As a future work, it would be interesting to consider whether our verification method can be improved by using the result in Ref.~\cite{PSSO20}.
At least, we can say that our construction seems to be more efficient than that using only the time-dividing technique in Ref.~\cite{PHOW19} in terms of the number of ancillary qubits.
As stated in Ref.~\cite{FN+}, when the target state is $|\psi_t\rangle=U|0^n\rangle$, the verification is possible by applying $U^\dag$ to $\hat{\rho}_{\rm out}$.
When we straightforwardly apply the time-dividing technique in Ref.~\cite{PHOW19} to $U^\dag$, in the worst case, $O(\log{n})$ ancillary qubits may be necessary because $D=O(\log{n})$, while our construction requires only two ancillary qubits.
In other words, although our method requires only $(m+1)$-qubit measurements, the direct use of Ref.~\cite{PHOW19} may require $(m+O(\log{n}))$-qubit ones.

A similar but different approach was also used in the Google team's supremacy experiment~\cite{google}.
To characterize their 53-qubit chip, they divided qubis on it into two small sets, which are called patches, by actually removing all interactions between them.
Their verification method was experimentally validated under their noise model.
However, their method would not work in the general case.
On the other hand, since our method ``virtually" removes some two-qubit gates (see Eq.~(\ref{decom})), it works in any case.

\section*{Acknowledgments}
We thank Samuele Ferracin for helpful discussions.
We also thank anonymous reviewers for insightful comments.
This work is supported by JST [Moonshot R\&D -- MILLENNIA Program] Grant Number JPMJMS2061 and the Grant-in-Aid for Scientific Research (A) No.JP22H00522 of JSPS.
Y. Takeuchi is supported by MEXT Quantum Leap Flagship Program (MEXT Q-LEAP) Grant Number JPMXS0118067394 and JPMXS0120319794.
TM is supported by the JST Moonshot R\&D JPMJMS2061-5-1-1, JST FOREST, MEXT Q-LEAP, the Grant-in-Aid for Scientific Research (B) No.JP19H04066 of JSPS, and the Grant-in-Aid for Transformative Research Areas (A) No.JP21H05183.
ST is supported by the Grant-in-Aid for Transformative Research Areas No.JP20H05966 of JSPS.

\bibliographystyle{quantum}

\begin{thebibliography}{9}
\bibitem{FN+}If $n$-qubit quantum operations are allowed, the efficient verification is trivially possible. Let $U$ be a unitary operator such that $|\psi_t\rangle=U|0^n\rangle$ for an ideal output state $|\psi_t\rangle$. We apply $U^\dag$ to a received state $\hat{\rho}$ and measure all qubits in the computational basis. Then, by estimating the probability of $0^n$ being observed, we can estimate the fidelity $\langle 0^n|U^\dag\hat{\rho}U|0^n\rangle$ between $|\psi_t\rangle$ and $\hat{\rho}$.
\bibitem{FN}For clarity, we use the notation $\hat{a}$ when the lowercase letter $a$ is a quantum state or quantum operation. On the other hand, for any uppercase letter $A$, we omit $\hat{\white{a}}$ even if $A$ is a quantum state or quantum operation.
\end{thebibliography}

\onecolumn\newpage
\appendix

\section{Estimation of the real part of Eq.~(\ref{cross})}
\label{A}
In this appendix, we show that the mean value of $\alpha$ converges to the real part of Eq.~(\ref{cross}).
Note that for simplicity, we write $\alpha({\bf i},{\bf j},{\bf i}',{\bf j}')$ by $\alpha$.
By using the spectral decomposition, we write $\hat{\rho}_{\rm out}=\sum_{i=1}^{2^n}p_i|\psi_i\rangle\langle\psi_i|$, where $p_i$ is a non-negative real number for any $i$, and $\sum_{i=1}^{2^n}p_i=1$. Therefore, the mean value $\mathbb{E}[\alpha \mid |0\rangle\langle 0|\otimes\hat{\rho}_{\rm out}]$ with the input state $|0\rangle\langle 0|\otimes\hat{\rho}_{\rm out}$ is written as $\sum_{i=1}^{2^n}p_i\mathbb{E}[\alpha \mid\ |0\rangle|\psi_i\rangle]$.

We explicitly calculate $\mathbb{E}[\alpha \mid\ |0\rangle|\psi_i\rangle]$.
The quantum state immediately before the measurements in Fig.~\ref{patch} (a) is
\begin{eqnarray*}
|\phi\rangle=\cfrac{|+\rangle\otimes Q_{{\bf i},{\bf j}}|\psi_i\rangle+|-\rangle\otimes Q_{{\bf i}',{\bf j}'}|\psi_i\rangle}{\sqrt{2}}
\end{eqnarray*}
with $|\pm\rangle\equiv(|0\rangle\pm|1\rangle)/\sqrt{2}$.
Remember that $\alpha=1$ if and only if $\oplus_{i=1}^nz_ik_i={\bf i}\cdot{\bf j}\oplus{\bf i}'\cdot{\bf j}'\oplus b$.
We separately consider the cases of $b=0$ and $b=1$.
The probability $P_1$ of obtaining $b=0$ and ${\bf z}$ such that $\oplus_{i=1}^nz_ik_i={\bf i}\cdot{\bf j}\oplus{\bf i}'\cdot{\bf j}'$ is
\begin{eqnarray*}
P_1=\langle\phi|\left(|0\rangle\langle 0|\otimes\cfrac{I^{\otimes n}+(-1)^{{\bf i}\cdot{\bf j}+{\bf i}'\cdot{\bf j}'}\prod_{i=1}^nZ_i^{k_i}}{2}\right)|\phi\rangle.
\end{eqnarray*}
On the other hand, the probability $P_2$ of obtaining $b=1$ and ${\bf z}$ such that $\oplus_{i=1}^nz_ik_i\neq{\bf i}\cdot{\bf j}\oplus{\bf i}'\cdot{\bf j}'$ is
\begin{eqnarray*}
P_2=\langle\phi|\left(|1\rangle\langle 1|\otimes\cfrac{I^{\otimes n}-(-1)^{{\bf i}\cdot{\bf j}+{\bf i}'\cdot{\bf j}'}\prod_{i=1}^nZ_i^{k_i}}{2}\right)|\phi\rangle.
\end{eqnarray*}
By a straightforward calculation,
\begin{eqnarray*}
P_1+P_2=\cfrac{1+(-1)^{{\bf i}\cdot{\bf j}+{\bf i}'\cdot{\bf j}'}{\rm Re}\left[\langle\psi_i|Q_{{\bf i},{\bf j}}^\dag \left(\prod_{i=1}^nZ_i^{k_i}\right)Q_{{\bf i}',{\bf j}'}|\psi_i\rangle\right]}{2},
\end{eqnarray*}
where ${\rm Re}[c]$ represents the real part of the complex number $c$.
Therefore,
\begin{eqnarray}
\nonumber
\mathbb{E}[\alpha\ \mid\ |0\rangle|\psi_i\rangle]&=&(P_1+P_2)\cdot1+(1-P_1-P_2)\cdot(-1)\\
\nonumber
&=&2(P_1+P_2)-1\\
\label{appa1}
&=&(-1)^{{\bf i}\cdot{\bf j}+{\bf i}'\cdot{\bf j}'}{\rm Re}\left[\langle\psi_i|Q_{{\bf i},{\bf j}}^\dag \left(\prod_{i=1}^nZ_i^{k_i}\right)Q_{{\bf i}',{\bf j}'}|\psi_i\rangle\right].\ \ \ \
\end{eqnarray}
From Eq.~(\ref{appa1}), the mean value $\mathbb{E}[\alpha \mid |0\rangle\langle 0|\otimes\hat{\rho}_{\rm out}]$ with the input state $|0\rangle\langle 0|\otimes\hat{\rho}_{\rm out}$ converges to
\begin{eqnarray*}
\sum_{i=1}^{2^n}p_i\mathbb{E}[\alpha \mid\ |0\rangle|\psi_i\rangle]=(-1)^{{\bf i}\cdot{\bf j}+{\bf i}'\cdot{\bf j}'}{\rm Re}\left[{\rm Tr}\left[\hat{\rho}_{\rm out}\left[Q_{{\bf i},{\bf j}}^\dag \left(\prod_{i=1}^nZ_i^{k_i}\right)Q_{{\bf i}',{\bf j}'}\right]\right]\right].
\end{eqnarray*}

\section{Conversion from Fig.~\ref{patch} (b) to (c)}
\label{B}
In this appendix, we show that $\sum_{{\bf l}\in\{0,1\}^3}\beta({\bf i},{\bf j},{\bf i}',{\bf j}',{\bf l})/2$ converges to the real part of Eq.~(\ref{cross}).
Let $U\equiv\Lambda(V_{{\bf i}'}^\dag)X\Lambda(V_{\bf i}^\dag)XH$ and $V\equiv H\Lambda(W_{{\bf j}'}^\dag)X\Lambda(W_{\bf j}^\dag)X$, where $\Lambda(Q)\equiv|0\rangle\langle0|\otimes I^{\otimes n'}+|1\rangle\langle1|\otimes Q$ for any $n'$-qubit unitary $Q$ and any natural number $n'$.
For simplicity, we define super-operators $\mathcal{U}(\cdot)\equiv U(\cdot)U^\dag$ and $\mathcal{V}(\cdot)\equiv V(\cdot)V^\dag$.
We also define $\Pi_{\alpha=1}$ and $\Pi_{\alpha=-1}$ as projectors onto the space satisfying $\alpha({\bf i},{\bf j},{\bf i}',{\bf j}')=1$ and $\alpha({\bf i},{\bf j},{\bf i}',{\bf j}')=-1$, respectively.
From them, we can define an observable $A\equiv\Pi_{\alpha=1}-\Pi_{\alpha=-1}$.

By using the above definitions, from Appendix~\ref{A}, the real part of Eq.~(\ref{cross}) can be written as
\begin{eqnarray}
\label{B1}
{\rm Tr}[A\mathcal{V}\mathcal{U}(|0\rangle\langle0|\otimes\hat{\rho}_{\rm out})].
\end{eqnarray}
Since any quantum state can be expanded by Pauli operators,
\begin{eqnarray}
\nonumber
\mathcal{U}(|0\rangle\langle0|\otimes\hat{\rho}_{\rm out})&=&(\mathcal{U}\otimes\mathcal{I}^{\otimes n-m})(|0\rangle\langle0|\otimes\hat{\rho}_{\rm out})\\
\label{B2}
&=&\sum_{i=1}^4\sum_{j=1}^{4^m}\sum_{k=1}^{4^{n-m}}c_{ijk}(\hat{\sigma}_i\otimes\hat{\tau}_j\otimes\hat{\chi}_k),\ \ \ \ \ \
\end{eqnarray}
where $\mathcal{I}(\cdot)=I(\cdot)I$, $c_{ijk}$ is a real number, and $\hat{\sigma}_i$, $\hat{\tau}_j$, and $\hat{\chi}_k$ are $1$-, $m$-, and $(n-m)$-qubit Pauli operators, respectively.
Furthermore, for any $\hat{\sigma}_i$, we utilize the well-known equality
\begin{eqnarray}
\label{B3}
\hat{\sigma}_i=\cfrac{{\rm Tr}[\hat{\sigma}_i]I+{\rm Tr}[\hat{\sigma}_iX]X+{\rm Tr}[\hat{\sigma}_iY]Y+{\rm Tr}[\hat{\sigma}_iZ]Z}{2},
\end{eqnarray}
where $Y$ is the Pauli-$Y$ gate.
From Eqs.~(\ref{B2}) and (\ref{B3}), Eq.~(\ref{B1}) is rewritten as
\begin{eqnarray*}
\sum_{i=1}^4\sum_{j=1}^{4^m}\sum_{k=1}^{4^{n-m}}\cfrac{c_{ijk}}{2}&\{&{\rm Tr}[\hat{\sigma}_i\otimes A[\hat{\tau}_j\otimes\mathcal{V}(|0\rangle\langle 0|\otimes\hat{\chi}_k)]]+{\rm Tr}[\hat{\sigma}_i\otimes A[\hat{\tau}_j\otimes\mathcal{V}(|1\rangle\langle 1|\otimes\hat{\chi}_k)]]\\
&&+{\rm Tr}[X\hat{\sigma}_i\otimes A[\hat{\tau}_j\otimes\mathcal{V}(|+\rangle\langle +|\otimes\hat{\chi}_k)]]-{\rm Tr}[X\hat{\sigma}_i\otimes A[\hat{\tau}_j\otimes\mathcal{V}(|-\rangle\langle -|\otimes\hat{\chi}_k)]]\\
&&+{\rm Tr}[Y\hat{\sigma}_i\otimes A[\hat{\tau}_j\otimes\mathcal{V}(|+i\rangle\langle +i|\otimes\hat{\chi}_k)]]-{\rm Tr}[Y\hat{\sigma}_i\otimes A[\hat{\tau}_j\otimes\mathcal{V}(|-i\rangle\langle -i|\otimes\hat{\chi}_k)]]\\
&&+{\rm Tr}[Z\hat{\sigma}_i\otimes A[\hat{\tau}_j\otimes\mathcal{V}(|0\rangle\langle 0|\otimes\hat{\chi}_k)]]-{\rm Tr}[Z\hat{\sigma}_i\otimes A[\hat{\tau}_j\otimes\mathcal{V}(|1\rangle\langle 1|\otimes\hat{\chi}_k)]]\},
\end{eqnarray*}
where $|\pm\rangle\equiv(|0\rangle\pm|1\rangle)/\sqrt{2}$ and $|\pm i\rangle\equiv(|0\rangle\pm i|1\rangle)/\sqrt{2}$.

By using Eq.~(\ref{clifford}), the above formula can be concisely represented by
\begin{eqnarray*}
&&\sum_{{\bf l}=000,001}\cfrac{{\rm Tr}[(I\otimes A)[\mathcal{U}\otimes\mathcal{V}'(|0\rangle\langle0|\otimes\hat{\rho}_{\rm out}\otimes C_{\bf l}|0\rangle\langle0|C_{\bf l}^\dag)]]}{2}\\
&&+\sum_{{\bf l}\neq000,001}\cfrac{{\rm Tr}[(C_{\bf l}ZC_{\bf l}^\dag\otimes A)[\mathcal{U}\otimes\mathcal{V}'(|0\rangle\langle0|\otimes\hat{\rho}_{\rm out}\otimes C_{\bf l}|0\rangle\langle0|C_{\bf l}^\dag)]]}{2}\\
&=&\cfrac{1}{2}\sum_{{\bf l}\in\{0,1\}^3}\mathbb{E}\left[\beta({\bf i},{\bf j},{\bf i}',{\bf j}',{\bf l}) \mid\ |0\rangle\langle0|\otimes\hat{\rho}_{\rm out}\otimes|0\rangle\langle0|\right],
\end{eqnarray*}
where $\mathcal{V}'\equiv V'(\cdot){V'}^\dag$, and $V'=U_{\rm perm}^\dag VU_{\rm perm}$ with the permutation gate $U_{\rm perm}$ shifting the $i$th qubit to the $(i+1\ ({\rm mod}\ n-m+1))$th one for $1\le i\le n-m+1$ (see also Fig.~\ref{patch} (c)).
Note that the term $(-1)^{(1+\delta_{l_1,0}\delta_{l_2,0})o}$ in the definition of $\beta({\bf i},{\bf j},{\bf i}',{\bf j}',{\bf l})$ means that when ${\bf l}=000$ or $001$, we ignore the measurement outcome $o$.
This concludes that $\sum_{{\bf l}\in\{0,1\}^3}\beta({\bf i},{\bf j},{\bf i}',{\bf j}',{\bf l})/2$ converges to the real part of Eq.~(\ref{cross}).

\begin{figure*}[t]
\centering
\includegraphics[width=13cm, clip]{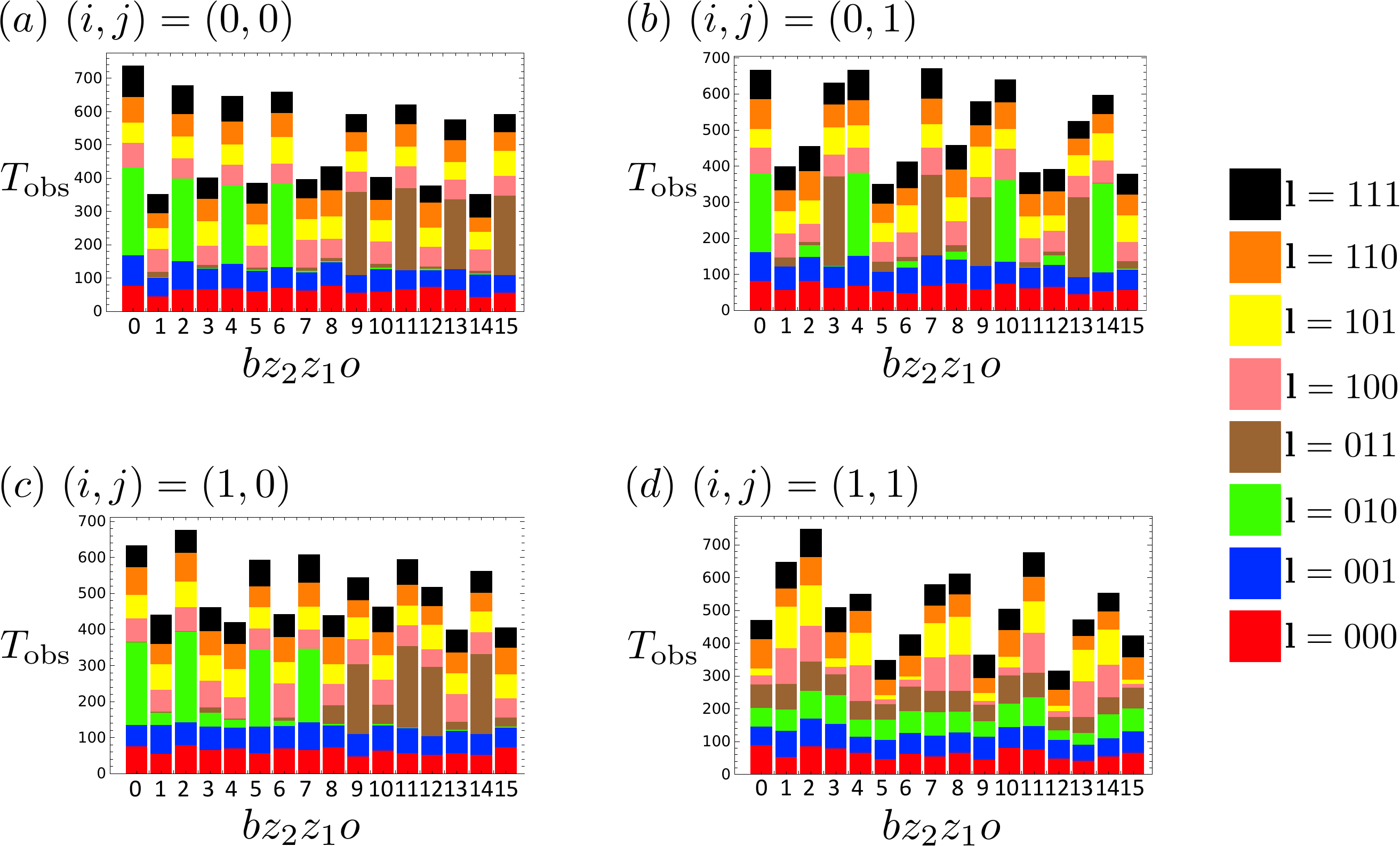}
\caption{Experimental results of our method performed on the IBM Manila 5-qubit chip. $T_{\rm obs}$ counts how many times each of $bz_2z_1o$ is observed. The labels of the horizontal axis denote the decimal numbers of $bz_2z_1o$, i.e., the values of $8b+4z_2+2z_1+o$. Note that since the circuits in the case of ${\bf l}=000$ and ${\bf l}=110$ are the same, we obtain $\beta'_k(i,j,0,0,000)$ and $\beta'_k(i,j,0,0,110)$ from the same experimental data.
The same applies to the case of ${\bf l}=001$ and ${\bf l}=111$.}
\label{experiment1}
\end{figure*}

\begin{figure}[t]
\centering
\includegraphics[width=9cm, clip]{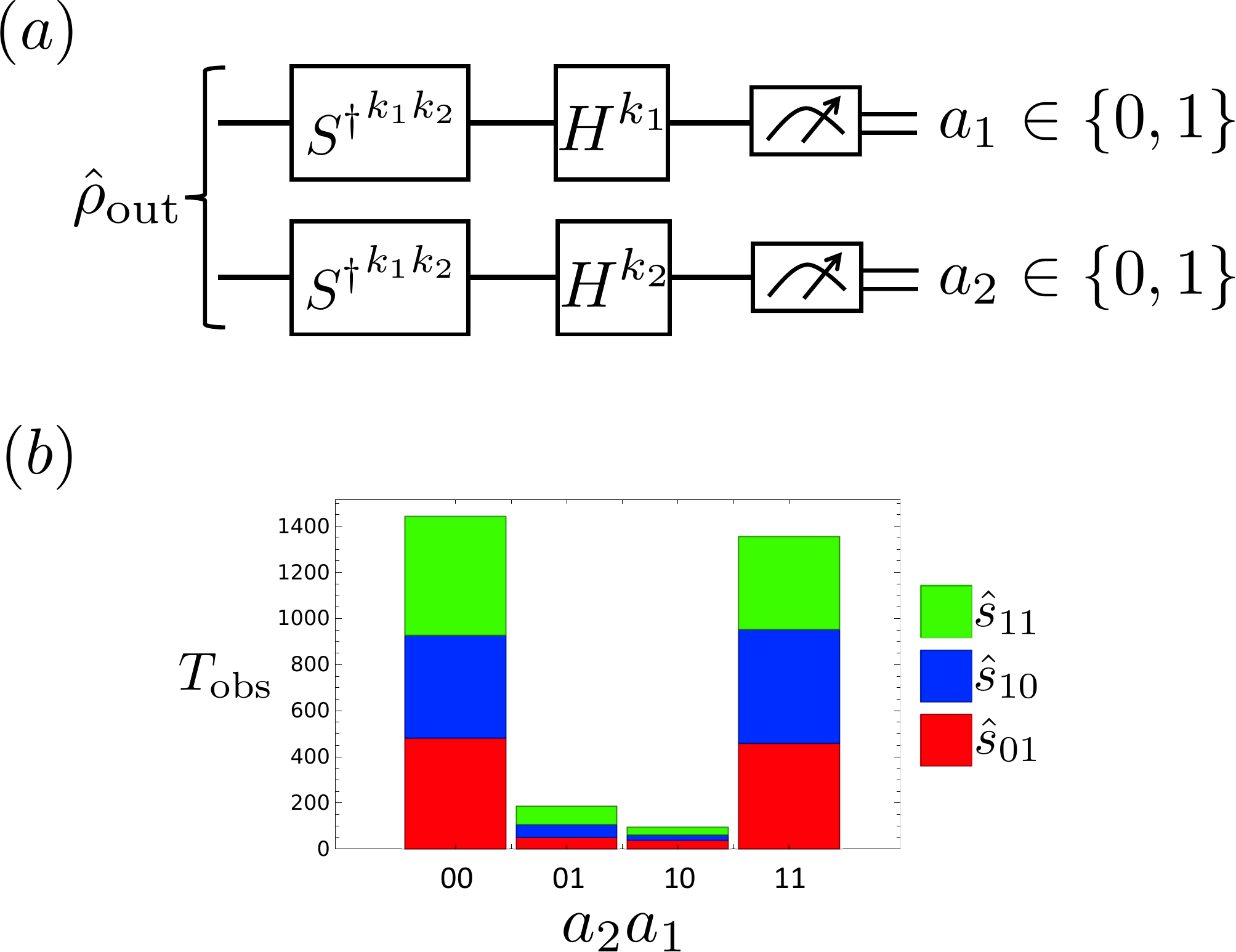}
\caption{Quantum circuits used to directly estimate ${\rm Re}[{\rm Tr}[\hat{\rho}_{\rm out}\hat{s}_{\bf k}]]$ for ${\bf k}\in\{01,10,11\}$ and their measurement outcomes. (a) $S^\dag\equiv|0\rangle\langle 0|-i|1\rangle\langle 1|$, and $k_1$ and $k_2$ are the 1st and 2nd bits of ${\bf k}$, respectively. Let $\gamma(a_1,a_2)\in\{1,-1\}$ be a random variable such that $\gamma(a_1,a_2)=1$ if and only if $a_1=a_2$. Since the expected value of $\gamma(a_1,a_2)$ is ${\rm Re}[{\rm Tr}[\hat{\rho}_{\rm out}\hat{s}_{\bf k}]]$, it can be estimated from $a_1$ and $a_2$. (b) We run the quantum circuit in (a) $1024$ times for each ${\bf k}\in\{01,10,11\}$. As a result, we obtain this histogram. $T_{\rm obs}$ counts how many times each of $a_2a_1$ is observed.}
\label{experiment2}
\end{figure}

\section{Experimental data}
\label{AC}
In this appendix, we give the data obtained from our experiment on the IBM Manila 5-qubit chip.
This chip is represented by a one-dimensional graph with five vertices.
We have used four qubits labeled by $Q0$, $Q1$, $Q2$, and $Q3$ and have performed the experiment in 30th April 2022.

For all $i,j\in\{0,1\}$ and ${\bf l}\in\{0,1\}^3$, we run simplified quantum circuits of those in Fig.~\ref{patch} (c) and obtain measurement outcomes $(b,z_2,z_1,o)\in\{0,1\}^4$, where $z_1$ and $z_2$ are the 1st and 2nd bits of ${\bf z}$, respectively.
The experimental results are summarized in Fig.~\ref{experiment1}.
For comparison, we also directly estimate ${\rm Re}[{\rm Tr}[\hat{\rho}_{\rm out}\hat{s}_{\bf k}]]$ for ${\bf k}\in\{01,10,11\}$.
The experimental results and quantum circuits used to obtain them are given in Figs.~\ref{experiment2} (b) and (a), respectively.

\end{document}